\documentclass[preprint,12pt]{aastex}
\usepackage{natbib}

\newcommand\dunits{h^3\;{\rm Mpc}^{-3}\;{\rm mag}^{-1}}
\newcommand\Rab{R_{\rm AB}}

\shorttitle{Lyman Break Galaxy Candidates in SDSS DR1}
\shortauthors{Bentz, Osmer, \& Weinberg}

\begin{document}
\title{Bright Lyman Break Galaxy Candidates in the Sloan Digital Sky Survey 
First Data Release}
\author{Misty C. Bentz, Patrick S. Osmer, and David H. Weinberg}
\affil{Department of Astronomy, The Ohio State University}
\affil{140 W. 18th Ave, Columbus, OH 43210-1173}
\email{bentz,posmer,dhw@astronomy.ohio-state.edu}

\begin{abstract}
We report the discovery of six compact, starburst galaxy candidates with
redshifts $2.3 < z < 2.8$ and $r$-band magnitudes 19.8$-$20.5 in the
Quasar Catalog of the SDSS DR1.  The SDSS spectra resemble the composite
spectrum of Lyman Break Galaxies (LBGs) at $z \approx 3$ (albeit with
some differences and broader spectral lines), but the objects are $4-5$
magnitudes brighter than an ``$L_*$'' LBG.  These objects could be
extremely luminous LBGs, normal LBGs amplified by gravitational lensing,
or a rare class of BAL quasars.  In the first case, star formation rates
inferred from the UV continuum luminosities, with no correction for dust
extinction, are $\sim$ 300$-$1000 ${\rm M}_{\odot}$ yr$^{-1}$, similar
to those of ultraluminous infrared galaxies, but in these UV-bright
objects the star formation is evidently not obscured by high dust column
densities.  The SDSS images show no evidence of multiple imaging or
foreground lensing structures.  The spectra have fairly broad absorption
features and prominent high-ionization absorption but do not show
high-ionization emission lines and do not resemble known BAL quasars.  A
rough estimate of the luminosity function of the objects is above a
Schechter function extrapolation from LBGs at fainter magnitudes.
Improved optical spectra and observations at X-ray, IR, and sub-mm
wavelengths could help determine the nature of these remarkable objects.
\end{abstract}

%\keywords{galaxies: starburst --- surveys}

\section{INTRODUCTION}
One of the most important recent developments in observational cosmology
is the discovery of a large population of star-forming galaxies at $z
\approx 3$ (e.g., \citealt{1996ApJ...462L..17S};
\citealt{1997ApJ...481..673L}).  These so-called Lyman break galaxies
(LBGs) can be identified by their characteristic colors in deep imaging
surveys.  Spectroscopic surveys of LBGs have transformed our
understanding of the history of cosmic star formation (e.g.,
\citealt{1996MNRAS.283.1388M}; \citealt{1999ApJ...519....1S}) and of
galaxy clustering in the early universe (e.g.,
\citealt{2003ApJ...584...45A}).  At $\Rab \approx 24.5$, the surface
density of LBGs is $\sim 1$ arcmin$^{-2}$, and the luminosity function
is well described by a \citet{1976ApJ...203..297S} function with a
characteristic luminosity corresponding to an R$_{AB}$ apparent
magnitude $m_* = 24.54$ at $z \approx 3 $ (Adelberger \& Steidel 2000).

Most known LBGs have been discovered in deep images covering small areas
of the sky.  The survey by \citet{2003ApJ...592..728S}, for example,
contains more than 2000 spectroscopically confirmed galaxies, but the 17
fields in the study cover only $\sim$ 0.38 deg$^2$.  The
brightest galaxies in this survey have $\Rab \approx 23$.  The only
substantially brighter object known, MS 1512-cB58 with $\Rab \approx
20.4$, was discovered serendipitously by \citet{1996AJ....111.1783Y} as
part of the CNOC-1 redshift survey, and is now understood to be
magnified by a factor $\sim 30$ as a result of gravitational lensing by
a foreground galaxy cluster \citep{1998MNRAS.298..945S}.  Detailed
studies of the stellar populations and interstellar gas properties of
LBGs rest either on this single object (e.g.,
\citealt{2000ApJ...533L..65T}; \citealt{2000ApJ...528...96P}) or on
composite spectra constructed from many individual galaxies
\citep{2003ApJ...588...65S}.  The limited area of existing LBG surveys
also means that the very bright end of the LBG luminosity function is
unconstrained.

The imaging survey of the Sloan Digital Sky Survey (SDSS,
\citealt{2000AJ....120.1579Y}) complements existing LBG imaging surveys
by covering a much wider area to a brighter limiting magnitude.
Furthermore, because the SDSS quasar selection algorithm
\citep{2002AJ....123.2945R} identifies unresolved sources with non-stellar
colors, rather than just objects with expected quasar colors, it {\it
also} selects the most luminous LBGs as quasar candidates and targets
them for spectroscopy.  In a search for quasars with anomalously low
\ion{C}{4} emission, \citet{paper1} found one example of an apparently
luminous, $z \approx 2.5$ starburst candidate that had been classified
as a quasar by the SDSS spectroscopic pipeline in the Early Data Release
(EDR, \citealt{2002AJ....123..485S}) Quasar Catalog
\citep{2002AJ....123..567S}.  In this {\it Letter},
we describe a detailed search through the SDSS First Data Release (DR1, \citealt{dr1preprint})
and the discovery of five additional objects with similar properties.
These objects have intriguing properties that necessitate a variety of
additional observations, and, if confirmed as starbursts, they would
provide the first constraints on the space density of extremely luminous
(in rest-frame UV) star-forming galaxies at high redshifts.

\section{SPECTRAL ANALYSIS}
As described by \citet{dr1qsopreprint}, the SDSS DR1 Quasar Catalog
covers $\sim$ 1360 $\rm{deg^2}$ of the sky and contains 16,713 objects
with rest-frame absolute magnitudes $M_i < -22$ (for $h \equiv H_0/ 100
\;{\rm km}\;{\rm s}^{-1}\;{\rm Mpc}^{-1} = 0.7$, $\Omega_M = 0.3$,
$\Omega_{\Lambda} = 0.7$, and a power law (frequency) continuum index of
$\alpha_Q=-0.5$), at least one emission line with a FWHM larger than
1000 km s$^{-1}$, and reliable redshifts.  For this investigation, we
searched the SDSS DR1 Quasar Catalog for high-$z$ starburst galaxies
that might have been classified as quasars, limiting the catalog to $z >
2.3$ in order to probe the high-$z$ regime of traditional LBG surveys.
Color cuts were then applied to the 1658 objects satisfying the redshift
cut.  Although the filter systems are different, those objects with
optical colors satisfying
\begin{equation}
g - r \leq 1.2,\ u - g \geq g - r + 1 
\end{equation}
are analogous to the objects that would be detected by the LBG color
criteria of \cite{2003ApJ...588...65S}.  A total of 591 objects met the
color criteria, and they were all visually inspected for absorption line
features similar to those in the LBG composite spectrum of
\citet{2003ApJ...588...65S}, and for the absence of characteristic
quasar emission lines from highly ionized species.  Six objects
(including the original object discovered in the EDR by
\citealt{paper1}) displayed these spectral signatures of a starburst
galaxy.  Table 1 lists the six candidate objects and their properties.
Figure 1 shows the individual spectra of the six candidates taken with
the 2.5$-$m SDSS telescope alongside the composite spectrum of 811
galaxies obtained with the 10$-$m Keck telescope by
\citet{2003ApJ...588...65S}.

The six candidates are red relative to normal quasars, but because they
have non-stellar colors and are unresolved at the resolution of SDSS
imaging, they meet the SDSS selection criteria for high-redshift quasar
candidates \citep{2002AJ....123.2945R} and were targeted for
spectroscopy. We checked the SDSS DR1 spectral database to confirm that
there were no objects classified as galaxies with $z > 2.3$ and no
high-$z$ quasar candidates satisfying our selection criteria that might
have been rejected from the Quasar Catalog because they failed to pass
the 1000 km s$^{-1}$ emission line cut.  

Table 1 lists the redshifts and broad-band photometric properties of the
six candidates, taken from the SDSS DR1 Quasar Catalog
\citep{dr1qsopreprint}.  All six objects are 4-5 magnitudes brighter
than an $L_*$ LBG, and 2-3 magnitudes brighter than the brightest LBG
detected by \citet{2003ApJ...592..728S}.

\section{LENSED LBGS, ULTRALUMINOUS LBGS, OR BAL QUASARS?}
From the absence of emission lines of highly ionized species, such as
\ion{N}{5} and \ion{C}{4}, it is obvious that these objects are not
normal AGN.  But are they normal LBGs that have been amplified by
gravitational lensing, extremely luminous LBGs, or unusual broad
absorption line (BAL) quasars?

The objects were found by searching for absorption and Ly$\alpha$
features similar to those of the LBG composite, so by definition they share
many spectral similarities.  There is strong evidence in most of the
candidates for absorption from \ion{Si}{2} $\lambda$1260, \ion{C}{2}
$\lambda$1334, \ion{Si}{4} $\lambda\lambda$ 1393,1402, \ion{C}{4}
$\lambda\lambda$1548, 1550, and \ion{Al}{3} $\lambda\lambda$1854, 1862.
In a few of the candidates, there is even noticeable absorption from
\ion{O}{1} $\lambda$1302, \ion{Si}{2} $\lambda$1304, \ion{Si}{2}
$\lambda$1526, \ion{Fe}{2} $\lambda$1608, and \ion{Al}{2} $\lambda$1670.
The precedent of MS 1512-cB58 suggests that the high apparent brightness
of these objects could be a consequence of lensing.  We inspected the
SDSS images of the six candidates for evidence of multiple images or
possible foreground galaxies or clusters, but all of the systems are
consistent with lone, unresolved point sources with unexceptional
foreground environments.  We also counted the numbers of objects
classified as extended in the DR1 imaging catalog within a 5\arcmin\
radius of five of the six objects,\footnote{We were not able to carry
out the counts for J1553+0056 because of a difficulty with the
database.} and similarly for control fields offset by $1\degr - 4\degr$.
The counts did not show any statistically significant excess of extended
sources around the candidates, except possibly for SDSS J1444+0134.  The
SDSS data are not sufficient to rule out sub-arcsecond image splitting
or the presence of foreground systems at likely lens redshifts of $z
\sim 1$.

At present, the main evidence against the lensing hypothesis comes from
the differences between the LBG composite spectrum and the spectra of
the galaxy candidates.  The candidates are redder than most LBGs, and
most of their absorption features are broader, as are their Ly$\alpha$
profiles.\footnote{The FWHMs of the Ly$\alpha$ emission lines range
from $\sim 1100-3000$ km s$^{-1}$.  SDSS J1340+6344 and SDSS J1432-0001
have fairly symmetric Ly$\alpha$ profiles, while the LBG composite has a
P-Cygni type asymmetry suggestive of outflows.}  While we would not
expect such differences if these were normal LBGs, they might arise if
the objects have extremely high star formation rates (SFRs).  If we
assume that lensing and AGN contributions are unimportant, we can follow
the standard practice of estimating SFRs from the continuum luminosity
at $\lambda$1500 \AA, which is produced mainly by young O and B stars.
We estimate the rest-frame $\lambda$1500
\AA\ flux directly from the calibrated spectra and convert to luminosity
assuming a cosmological model with $\Omega_M=0.3$, $\Omega_\Lambda=0.7$,
and $h=0.7$.  Assuming a Salpeter IMF with mass limits of 0.1 to 100
M$_{\odot}$ and a 10$^8$ yr old continuous star formation model,
\citet{Kennicutt1998} derives the relation
\begin{equation}
\rm SFR\ M_{\odot}\ yr^{-1} = 1.4 \times 10^{-28} L_{\nu}(\lambda 1500)
\end{equation} 
between the star formation rate and the luminosity density $L_\nu$ (in
ergs s$^{-1}$ Hz$^{-1}$) at $\lambda=1500$\AA.  The implied SFRs are
tabulated in Table 2, along with the measured flux in the observed frame
at $\lambda$1500 \AA\ $\times (1+z)$.  As expected for objects as bright
as these, the SFRs are very high, ranging from $\sim$ 300 M$_{\odot}$
yr$^{-1}$ to over 1000 M$_{\odot}$ yr$^{-1}$.  

The values in Table~2 assume no dust extinction and are therefore lower
limits to the ``true'' SFRs, at least if the assumed IMF is correct.
Lyman break galaxies in typical spectroscopic surveys have similarly
estimated SFRs in the range $\sim 50-100 h_{70}^{-2}\;{\rm
M}_\odot\;{\rm yr}^{-1}$ {\it after} correction for an average factor of
$\sim$ 7 attenuation by dust \citep{2003ApJ...588...65S}.  As previously
mentioned, our candidates are redder than the LBGs in the
\citet{2003ApJ...592..728S} survey, which suggests that their extinction
corrections should be larger, pushing the corrected SFRs to thousands of
M$_\odot\;{\rm yr}^{-1}$.  The stellar winds and interstellar turbulence
associated with such extreme SFRs could account for the relatively broad
absorption features and the strength of high excitation absorption seen
in some of our candidates.

The brightest sub-mm galaxies are inferred to have comparable or, in
some cases, even higher SFRs at these redshifts (see, e.g.,
\citealt{2003Natur.422..695C}), but in these objects most of the energy
from massive young stars is absorbed and re-radiated by dust.  The SCUBA
source in N2 850.4 \citep{Smail2003}, with an estimated SFR $\sim 300$
M$_{\odot}$ yr$^{-1}$, has an optical spectrum reminiscent of those in
Figure~1, though our objects do not show P-Cygni profiles.
High-redshift galaxies with such SFRs {\it should} appear in the SDSS if
they have moderate UV extinction, and our candidates could plausibly
represent a later evolutionary stage of luminous sub-mm galaxies, with
young stellar populations that have burned through enough of their dusty
envelopes to become visible in the rest-frame UV.

Finally, there is the possibility that these luminous objects are
powered by black hole accretion rather than star formation, and that
they represent a very unusual class of BAL quasars.  In favor of this
interpretation, two of the spectra show evidence for broad \ion{C}{3}]
emission, and some have very broad absorption lines, up to $\sim 7600$
km s$^{-1}$.  The strengths of the absorption features from highly
ionized species are also much stronger than the low ionization
absorption features.  However, the candidates do not show typical BAL
profiles, and we are not aware of any analogs of these objects among
previously discovered BALs.  There is significant variation among the
spectra in Figure~1, and the relative importance of star formation and
AGN activity could be different from object to object.

\section{THE BRIGHT END OF THE LBG LUMINOSITY FUNCTION}
If we assume that the ultraluminous LBG interpretation is correct, and
that we can therefore ignore lensing amplification and AGN
contributions, then we can constrain the bright end of the rest-frame UV
luminosity function of high-redshift galaxies.  Our estimate here is
necessarily crude, (1) because we have only six objects, and (2) because
assessing completeness as a function of luminosity and redshift would
require detailed simulation of the SDSS quasar target selection
algorithm.  We make the conservative assumption that the selection
efficiency is unity in the redshift range $2.3 < z < 2.8$ over the 1360
deg$^2$ area of DR1, for objects above the $i=20.2$ apparent magnitude
limit of the quasar catalog.  Our objects lie within a 1-magnitude bin,
with median redshift $z = 2.55$ and median absolute magnitude
$M_{\rm UV} \approx -25.4$.\footnote{We calculate the absolute magnitude
from the $i$-band apparent magnitude, so it corresponds to rest-frame
wavelength 7600\AA/$(1+z) \approx 2100$\AA.}  We estimate $\Phi(M) =
\sum_{j=1}^6 1/V_{a,j} = 2.52 \times 10^{-9}
\dunits$, where $V_{a,j}$ is the comoving accessible volume for galaxy
$j$, which we take to run from $z=2.3$ to the smaller of $z=2.8$ and
$z_{\rm max}$, the redshift at which the galaxy's apparent magnitude
would hit the $i=20.2$ limit.  The $z=2.8$ upper limit is based on the
maximum redshift of our candidates, and while the selection function may
continue beyond this redshift, it is likely to decline as Ly$\alpha$
moves past the peak of the $g$-filter.  If we assumed instead that each
galaxy could be seen all the way to $z_{\rm max}$, then our density
estimate would drop by 20\%.

Figure 2 shows this estimate in the context of the LBG luminosity
function estimated by \citet{2000ApJ...544..218A} from a combination of
ground-based data and the Hubble Deep Field.  Circles show the
\citet{2000ApJ...544..218A} data points, with no correction for dust
extinction, and the solid line shows their Schechter function fit.  The
open triangle shows our above estimate at $z = 2.55$, with an error bar
that corresponds to the Poisson error on six objects.  We plot our point
at the median apparent magnitude $r \approx 20.3$, ignoring the
difference between the Sloan $r$-band and Adelberger \& Steidel's
$R$-band.  Our assumption of unit selection efficiency is probably too
generous, so the most likely value for the space density is above the
data point.  Since the
\citet{2000ApJ...544..218A} data are given in terms of apparent
magnitude and their galaxies have a median $z \approx 3$, we also plot a
filled triangle that shows the effect of moving our galaxies from
$z=2.55$ to $z=3$, using a $K$-correction based on the median spectral
slope of $\alpha = -2.16$, with $f_{\nu} \propto \nu^{\alpha}$, derived
by fitting the $griz$ magnitudes.

Our estimate of the luminosity function at $r \approx 20.3$ is above the
Schechter function extrapolation from fainter magnitudes.  However,
there is no reason to expect that the high-redshift, rest-frame UV
luminosity function should be described by a Schechter function over
such a wide range in luminosity, especially given the importance of
recent star formation history and the geometry of extinction in
determining a galaxy's luminosity at $\lambda 1500$\AA.  A power law
connecting our data point (offset to $z=3$) to the brightest data point
from \citet{2000ApJ...544..218A} has a slope of $\Rab^{-1.93}$ (or
$L^{-2.93}$).

A variety of follow-up observations could help to determine the nature
of these remarkable objects.  High-resolution imaging to search for
image splitting and deeper imaging to search for foreground structures
could test the lensing hypothesis.  Better optical spectra would allow
better measurements of absorption profiles and tighter constraints on
broad emission lines, and they might reveal high-excitation photospheric
lines that would be the signature of starburst activity.  Near-IR
spectroscopy, sub-mm imaging, and X-ray measurements could better
quantify the importance of dust extinction and AGN activity.  If these
objects are confirmed as extreme LBGs, they demonstrate that the
stupendous starbursts so often veiled by thick layers of dust can
sometimes emerge to show their true colors.

\acknowledgements
We would like to thank Patrick Hall, Ian Smail, Charles Steidel, and an
anonymous referee for helpful comments and conversations.  Misty Bentz
is supported by a Graduate Fellowship of The Ohio State University.

The discovery of $z=2.5$ galaxies by a 2.5-m telescope is a tribute to
the power of a large area, multi-color imaging survey, to the efficiency
of the SDSS spectrographs and spectroscopic pipeline, and to the utility
of the SDSS Archive.  The SDSS Web Site, http://www.sdss.org , lists the
Participating Institutions and the project's other sources of funding,
which include the Sloan Foundation, the NSF, NASA, DOE, the Japanese
Monbukagakusho, and the Max Planck Society.  

\bibliographystyle{apj}
\bibliography{ms}

\clearpage
\begin{deluxetable}{lcccccccc} 
\tabletypesize{\footnotesize}
\tablecolumns{9} 
\tablewidth{0pc}
\tablecaption{Properties of Starburst Candidates}
\tablehead{ 
\colhead{} & 
\colhead{} &
\multicolumn{7}{c}{Colors and Magnitudes\tablenotemark{a}}\\ 
\cline{3-9}\\
\colhead{Object} & 
\colhead{z\tablenotemark{a}} & 
\colhead{u$-$g} &
\colhead{g$-$r} &
\colhead{r$-$i} &
\colhead{i$-$z} &
\colhead{r Obs.} &
\colhead{i Obs.} & 
\colhead{i Abs.\tablenotemark{b}}}
\startdata
SDSS J024343.77$-$082109.9	& 2.59 & 2.181 & 0.390 & 0.319 & 0.500 & 20.403 & 20.084 & -25.916 \\
SDSS J114756.00$-$025023.5\tablenotemark{c}	& 2.56 & 3.629 & 1.170 & 0.532 & 0.679 & 19.825 & 19.293 & -26.678 \\ 
SDSS J134026.44+634433.2	& 2.79 & 2.453 & 0.689 & 0.465 & 0.340 & 19.823 & 19.358 & -26.792 \\
SDSS J143223.10$-$000116.4\tablenotemark{d}	& 2.47 & 1.885 & 0.615 & 0.399 & 0.609 & 20.514 & 20.115 & -25.810 \\
SDSS J144424.55+013457.0	& 2.66 & 2.405 & 0.622 & 0.433 & 0.349 & 20.510 & 20.077 & -26.028 \\
SDSS J155359.96+005641.3 	& 2.63 & 2.719 & 0.562 & 0.523 & 0.447 & 20.199 & 19.676 & -26.524 \\
\enddata
\tablenotetext{a} {Values taken from \citet{dr1qsopreprint}.}
\tablenotetext{b} {As determined by \citet{dr1qsopreprint}, with 
		$H_{0} = 70$ km s$^{-1}$ Mpc$^{-1}$, $\Omega_M = 0.3$,
		$\Omega_{\Lambda} = 0.7$; note that this absolute
		magnitude is extrapolated to rest-frame $i$-band
		assuming $\alpha_Q = -0.5$, while our objects are
		much redder.  }
\tablenotetext{c} {SDSS J114756.00$-$025023.5 is also a 2MASS 
		   object \citep{2003yCat.2246....0C}.}
\tablenotetext{d} {SDSS J143223.10$-$000116.4 is also an EDR
		   object (\citealt{2002AJ....123..567S};
		   \citealt{paper1}).}
\end{deluxetable}

\begin{deluxetable}{lcc}
\tabletypesize{\footnotesize}
\tablecolumns{3} 
\tablewidth{0pc}
\tablecaption{Star Formation Rates}
\tablehead{ 
\colhead{} & 
\colhead{F$_{\lambda}$($\lambda 1500 \times (1+z)$)\tablenotemark{a}} &
\colhead{SFR\tablenotemark{b}}\\
\colhead{Object} &
\colhead{(10$^{-17}$ erg s$^{-1}$ cm$^{-2}$ \AA$^{-1}$)} &
\colhead{(M$_{\odot}$ yr$^{-1}$)}}
\startdata
SDSS J0243$-$0821	& 2.7	& 550 \\
SDSS J1147$-$0250	& 2.7 	& 530 \\
SDSS J1340+6344		& 4.1	& 1100 \\	
SDSS J1432$-$0001	& 1.8 	& 320 \\	
SDSS J1444+0134		& 3.7	& 820 \\	
SDSS J1553+0056		& 2.6 	& 560 \\
\enddata
\tablenotetext{a}{Flux in observed frame at $\lambda = 1500 \times (1+z)$,
                  with a characteristic uncertainty of $0.3 \times
                  10^{-17}$ erg s$^{-1}$ cm$^{-2}$ \AA$^{-1}$ }
\tablenotetext{b}{Assumes a Salpeter IMF with mass limits between 0.1 
		  and 100 M$_{\odot}$ and a 10$^8$ year old continuous
		  star formation model \citep{Kennicutt1998}.  No
		  corrections for dust have been applied.}
\end{deluxetable}

\clearpage

\begin{figure}
\figurenum{1}
\epsscale{0.6}
\plotone{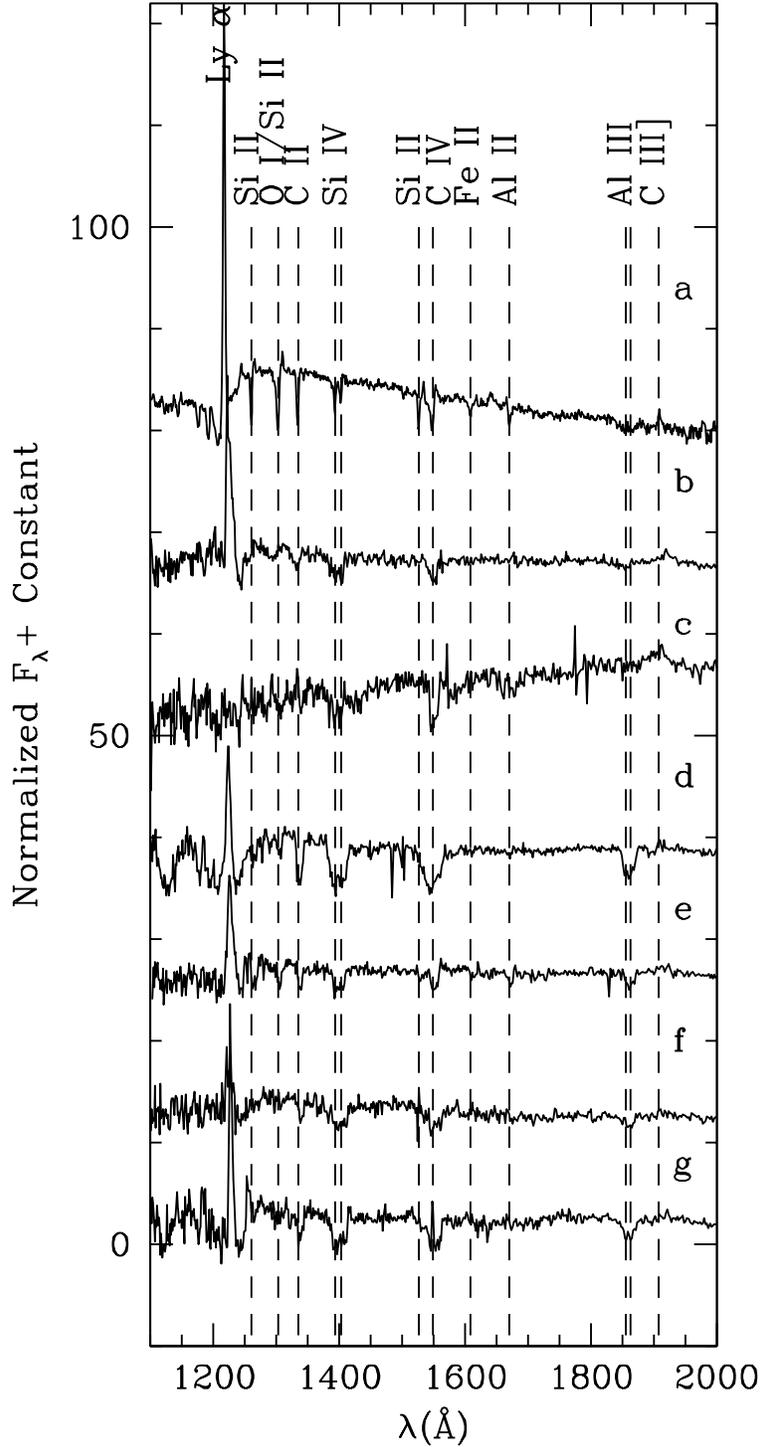}
\caption{Rest frame spectra of the Lyman break composite (Fig. 1a, 
         \citealt{2003ApJ...588...65S}) and the six candidate starburst
         galaxies. The Sloan spectra have been smoothed over five pixels
         and are as follows: b.) SDSS J0243-0821, c.) SDSS J1147-0250,
         d.) SDSS J1340+6344, e.) SDSS J1432-0001, f.) SDSS J1444+0134,
         and g.) SDSS J1553+0056} 
\end{figure} 

\begin{figure}
\figurenum{2} 
\epsscale{1} 
\plotone{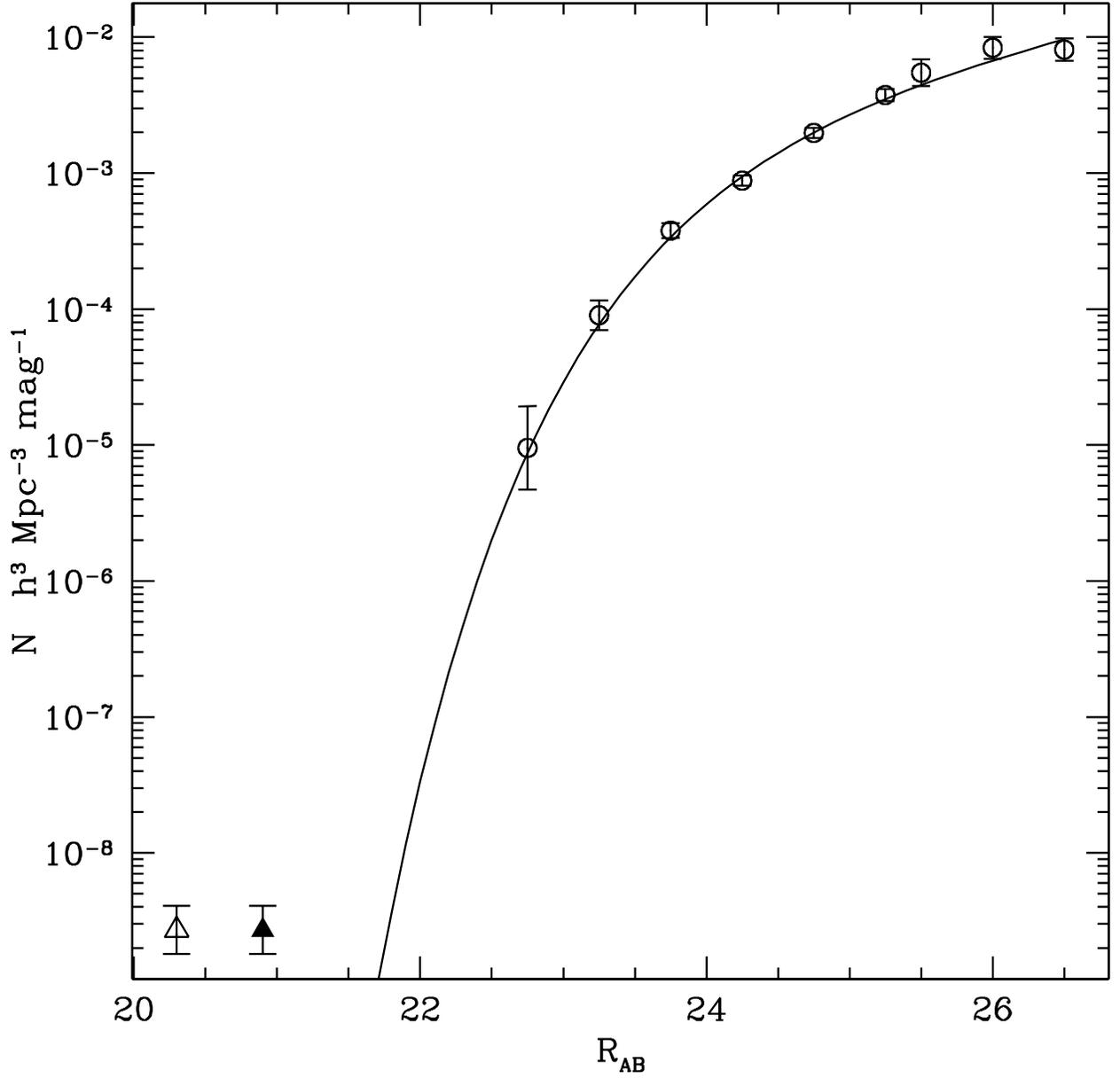}
\caption{Luminosity function of starburst galaxies at redshifts
         $z \approx$ 3 (for $\Omega_M$ = 0.3, $\Omega_{\Lambda}$ = 0.7).
         The circles are data points from \citet{2000ApJ...544..218A},
         and the solid curve shows their Schechter function fit with
         parameters $\Phi_* = 4.4\times 10^{-3} \dunits$, $m_* = 24.54$,
         $\alpha=-1.57$.  The open triangle is the additional data point
         based on the objects described in this work, with errorbars
         showing the Poisson noise for six objects.  The filled triangle
         shows the effect of changing from $z=2.55$, the center of our
         redshift bin, to $z=3$, the typical redshift of the
         \citet{2000ApJ...544..218A} objects.}
\end{figure}

\end{document}